\documentstyle[psfig,sprocl]{article}

\def\Journal#1#2#3#4{{#1} {\bf #2}, #3 (#4)}

\def\NPA{{\em Nucl. Phys.} A}
\def\PLB{{\em Phys. Lett.}  B}
\def\PRL{\em Phys. Rev. Lett.}
\def\PRC{{\em Phys. Rev.} C}
\def\be{\begin{equation}}
\def\ee{\end{equation}}
\def\bea{\begin{eqnarray}}
\def\eea{\end{eqnarray}}
\def\etal{{\it et al.}}

\begin{document}

\title{SUB-BARRIER FUSION OF DRIP-LINE NUCLEI}

\author{K. HAGINO}

\address{Institute for Nuclear Theory, Department of Physics, 
\\ University of Washington, Box 351550, Seattle, WA 98195, USA 
\\E-mail: hagino@phys.washington.edu}

\author{A. VITTURI}

\address{Dipartimento di Fisica, Universit\`a di Padova and INFN, 
   Padova, Italy\\E-mail: andrea.vitturi@pd.infn.it}


\maketitle
\abstracts{ 
We discuss the role of break-up process of a loosely-bound 
projectile in subbarrier fusion reactions. 
Coupled-channels calculations are carried out for 
$^{11}$Be + $^{208}$Pb and  $^{4,6}$He + $^{238}$U
reactions by discretizing in energy the particle continuum states. 
Our calculations show that the coupling to the break-up channel 
has two effects, namely the loss of flux and the dynamical modulation of 
fusion potential. Their net effects differ depending on 
the energy region. At energies above the Coulomb barrier, the former 
effect dominates over the latter and cross sections for complete fusion are 
hindered compared with the no coupling case. 
On the other hand, at below the barrier, the latter effect is much larger 
than the former and complete fusion cross sections are enhanced consequently. }

\section{Introduction}

Many questions 
concerning the effects of break-up process on subbarrier fusion
have been raised during the last few years 
both from the experimental~\cite{T97,R98,S98,D99,K98,T00} 
and theoretical~\cite{MPCD92,TKS93,DV94,Y97} points of view.  
The issue has become especially relevant in recent years due to the 
increasing availability of radioactive beams.  
These often involve weakly-bound systems close to the drip lines for
which the possibility of projectile dissociation prior to or at the 
point of contact cannot be ignored. 

One interesting question is whether the break-up process hinders or enhances 
fusion cross sections. In addressing this, it is vital to 
specify which quantity is being compared with which quantity; otherwise 
a comparison is meaningless. 
The latter quantity is rather obvious in theoretical calculations, 
because one can artificially switch on and off couplings 
to the break-up channel. 
When we refer in this paper to that break-up enhances/hinders fusion 
cross sections, it will be in this sense unless we mention explicitly. 
As for the 
former quantity, there are two possibilities, namely complete fusion cross 
sections or the sum of complete and incomplete fusion cross sections. 
From studies of fusion of stable nuclei where break-up process is not 
important, we have learned that any coupling of the 
relative motion of the colliding nuclei to nuclear intrinsic excitations 
causes large enhancements of the fusion cross section at subbarrier 
energies over the predictions of a simple barrier penetration 
model.~\cite{review} 
It may not be difficult to imagine that 
the same thing happens to the break-up channel as well; 
cross sections for 
inclusive processes, i.e., the sum of complete and incomplete fusion cross 
sections would be enhanced by  
couplings to the break-up channel. 
On the other hand, one could also argue 
intuitively that 
break-up process removes a part of flux and thus cross sections 
for complete fusion would be hindered. 
As will be shown below, 
fusion cross sections are determined by the 
competition of these two mechanisms. 

In passing, this sort of consideration is relevant also when one discusses 
experimental 
data and compares them with theoretical calculations. 
For instance, it is important to bear in mind that the recent 
Padova/RIKEN data~\cite{S98} as well as the Canberra data,~\cite{D99} which 
used $^9$Be beams, contain both the complete and incomplete fusion cross 
sections for the neutron break-up channel ($^9$Be $\to$ $n$ + $^8$Be), 
while they are only the complete fusion cross  sections 
with respect to the $\alpha$-break-up channel ($^9$Be $\to$ 
$\alpha$ + $^5$He or 2$\alpha$ + $n$). 

In the past, different theoretical approaches to the problem have led to 
controversial results, not only quantitatively but also qualitatively. 
Hussein \etal \,derived a local dynamical polarization potential 
$V_{\rm{DPP}}$ 
for break-up process and computed complete fusion cross sections 
as~\cite{MPCD92} 
\begin{equation}
\sigma_{\rm{CF}}=\frac{\pi}{k^2}\sum_l (2l+1)P_F^{(0)}
\exp\left(\frac{2}{\hbar}
\int^{\infty}_{-\infty}dt\,W(r(t))\right)\, ,
\end{equation}
where $P_F^{(0)}$ is the fusion probability {\it in the absence of break-up}, 
$k$ is the wave number in the entrance channel, and 
$W(r)$ is the imaginary part of the dynamical poralization potential 
$V_{{\rm DPP}}$. The integral in the exponent is evaluated along the classical 
trajectory $r(t)$, where $r(t=0)$ corresponds to the distance of 
closest approach. 
The idea of this formalism is that the effects of break-up can 
be well described in terms of the survival probability 
$P_{\rm{surv}}=1-P_{\rm{bu}}=\exp(2\int W(r)\,dt/\hbar )$. 
Since $W(r)$ is negative, the break-up process always hinders complete fusion 
cross sections in this formalism. Subsequently, Takigawa \etal \, pointed 
out that the classical trajectory used by Hussein \etal\, 
corresponded to the one for scatterings, and 
for fusion reactions the time integral should have been  
from $-\infty$ to 0.~\cite{TKS93} 
Consequently the effects of break-up were moderated over the estimate made by 
Hussein \etal, but the qualitative conclusion remained the same. 
These conclusions were later criticized by Dasso and Vitturi. They 
performed coupled-channels calculations by treating 
the break-up channels as a single state and obtained totally different 
results, i.e., enhancement of complete fusion cross sections due to 
the break-up.~\cite{DV94}

What would be the origin of these apparently controversial results? 
We argue that the coupling to the break-up channel leads to 
the dynamical modification of fusion potential as in fusion of stable 
nuclei. At subbarrier energies, this effect is most relevant, leading 
to the enhancement of complete fusion cross section. 
Such effects are automatically included in the coupled-channels formalism. 
The dynamical 
modulation of fusion potential is related to the real part of a 
dynamical polarization potential, which both Hussein \etal\, and 
Takigawa \etal\, completely threw away. 
Therefore, if the real 
part of the polarization potential is properly taken into account, 
their formalism would provide an enhancement of complete 
fusion cross sections at energies below the Coulomb barrier. 
Actually, 
Ref. 12 
shows that it is indeed the case, although their 
calculations are not satisfactory in a sense of Takigawa \etal~\cite{TKS93} 
and a full calculation within their formalism has still been awaited. 

Another origin of the controversy might be 
the simplified assumption used by Dasso and Vitturi.
As mentioned 
above, the entire continuum 
space was mocked up by a single discrete configuration,~\cite{DV94} 
and therefore the main feature of 
continuum couplings were not entirely included. 
This would underestimate the break-up effects, especially 
those effects, if any, which remove the incident flux. 

From this point of view, in this paper, we repeat similar calculations as 
of Dasso and Vitturi, but by considerably increasing the number of 
continuum states.~\cite{HV00}
We also use microscopic form factors in contrast to the previous 
studies, where form factors were assumed to have an exponential form. 
These two aspects enable us to describe simultaneously the dynamics of 
continuum couplings inside and outside the Coulomb barrier, leading to 
a more conclusive result on the effects of break-up process on subbarrier 
fusion. In the next section, we review briefly the coupled-channels 
formalism and define the complete and incomplete fusion cross sections. 
The specific applications of the formalism to 
$^{11}$Be + $^{208}$Pb and  $^{4,6}$He + $^{238}$U 
are given in the following section 3. 

\section{Coupled-Channels Formalism for Subbarrier Fusion of Drip-Line Nuclei} 

Our aim is to discuss effects of the break-up of the projectile nucleus 
on subbarrier fusion reactions by solving coupled-channels equations. 
The coupled-channels equations in the isocentrifugal approximation 
are given by~\cite{HRK99}
\begin{eqnarray}
&&\left[-\frac{\hbar^2}{2\mu}\frac{d^2}{dr^2}
+\frac{l(l+1)\hbar^2}{2\mu r^2}
+V_N^{(0)}(r)+\frac{Z_PZ_Te^2}{r}+\epsilon_n
-E\right]\psi_n(r) \nonumber \\
&&\qquad =-\sum_m\,F_{nm}(r)\psi_m(r),
\label{cc}
\end{eqnarray}
where the angular momentum of the relative motion in each channel 
has been replaced by the total angular momentum $l$.~\cite{HTBB95} 
In eq. (\ref{cc}), 
$V_N^{(0)}$ is the nuclear potential in the entrance channel, and 
$\epsilon_n$ is the excitation energy of the $n$-th channel.
Here, we assume that $n=0$ labels the ground state of the projectile 
nucleus and the all other $n$ refer to particle continuum states, which 
are associated with the break-up channels. 
It is straightforward to include bound excited states of the projectile 
and/or the excitations in the target nucleus. 
$F_{nm}$ are the coupling form factors, 
which we compute on the microscopic basis. 
It is to fold 
the external nuclear and Coulomb fields with the
proper single-particle transition densities, 
obtained by promoting the
last weakly-bound nucleon to the continuum states.
More explicitely, the single-particle form factor for the promotion from
the bound state $(n_1\ell_1j_1m_1)$ to the continuum state 
$(\ell_2j_2m_2)$ with continuum energy $E$ assumes the form
\begin{eqnarray}
&&   F_{n_1\ell_1j_1m_1\to E\ell_2j_2m_2} (r) =
\nonumber \\
&&   =  \sqrt{\pi}~\sum_{\lambda} 
     (-1)^{m_2+{1 \over 2}}~
 \delta(\ell_1+\ell_2+\lambda,{\rm even})~
  \left\langle j_1 {1 \over 2} j_2 -{1 \over 2} | \lambda 0 \right\rangle~ 
\nonumber \\
&& \times~
 { {\sqrt{2j_1+1}\sqrt{2j_2+1}} \over {\sqrt{2\lambda+1}} }~
 \langle j_1 -m_1 j_2 m_2 | L (m_2-m_1) \rangle~
\sqrt{\frac{2\lambda+1}{4\pi}}\,\delta_{m_1,m_2}
\nonumber \\
&& \times\left[ \int_{0}^{\infty}r'^2 dr' \int_{-1}^{+1}du\,
    R^*_{E\ell_2j_2} (r') R_{n_1\ell_1j_1} (r')
      V_T \big ( \sqrt{r^2+r'^2-2rr'u} \big ) P_\lambda (u) \right ]~.
\label{ff}
\end{eqnarray}
The functions  $R_{E\ell_2j_2} (r)$ and $ R_{n_1\ell_1j_1} (r)$ are the
single particle wave functions for the continuum and bound states,
respectively.  The potential $V_T$ is the mean field felt by the single 
particle due to the presence of the target, responsible of the transition, 
not to be confused 
therefore with the projectile mean field generating the single particle 
wave functions.  This potential obviuosly involves both a nuclear and 
a Coulomb component.
Note that for long ranged transition densities, 
as in the case of a weakly bound 
system, the resulting Coulomb formfactor will differ from the pure
$r^{-\lambda-1}$ form much outwards than in the traditional case of stable 
systems. See Ref. [16]
for more details. 

The coupled-channels equations (\ref{cc}) are solved by imposing the incoming 
wave boundary condition (IWBC), where there are only incoming waves 
at $r_{{min}}$ which is taken somewhere inside the Coulomb 
barrier.~\cite{HRK99,LP84}
The boundary conditions are thus expressed as
\begin{eqnarray}
\psi_n(r)&\to& T_n\exp\left(-i\int^r_{r_{min}}k_n(r')dr'\right)
~~~~~~~~~~~~~~~r \le r_{min}, \label{bc1}\\
&\to&
H_l^{(-)}(k_nr) \delta_{n,0} + R_nH_l^{(+)}(k_nr)~~~~~~~~~~~~~r>r_{max},
\label{bc2}
\end{eqnarray}
where
\begin{equation}
k_n(r)=\sqrt{\frac{2\mu}{\hbar^2}\left(E-\epsilon_n
-\frac{l(l+1)\hbar^2}{2\mu r^2}
-V_N^{(0)}(r)-\frac{Z_PZ_Te^2}{r}\right)}
\end{equation}
is the local wave number for the $n$-th channel and
$k_n=\sqrt{2\mu(E-\epsilon_n)/\hbar^2}$.
$H_l^{(-)}$ and $H_l^{(+)}$ are the incoming and the
outgoing Coulomb functions, respectively.

The fusion probability is defined as the ratio between the flux inside the 
Coulomb barrier and the incident flux. 
For our boundary conditions given by 
eqs. (\ref{bc1}) and (\ref{bc2}), it reads 
\begin{equation}
P_n=\frac{k_n(r_{min})}{k_0}\left|T_n\right|^2
\end{equation}
for the $n$-th channel. 
Complete fusion is a process where all the nucleons of the projectile 
are captured by the target nucleus. 
We thus define cross sections of complete fusion using 
the flux for the non-continuum channel (i.e., $n$=0) as~\cite{HT98} 
\begin{equation}
\sigma_{{\rm CF}} = \frac{\pi}{k_0^2}\sum_l (2l+1)P_0
= \frac{\pi}{k_0^2}\sum_l (2l+1)\,\frac{k_0(r_{min})}{k_0}\left|T_0\right|^2. 
\end{equation}
The flux for the particle continuum channel ($n\neq0$) are associated with 
incomplete fusion, whose cross sections we define as 
\begin{equation}
\sigma_{{\rm ICF}} = \frac{\pi}{k_0^2}\sum_l (2l+1)\,\sum_{n\neq 0}P_n
= \frac{\pi}{k_0^2}\sum_l (2l+1)\,
\sum_{n\neq 0}\frac{k_n(r_{min})}{k_0}\left|T_n\right|^2. 
\end{equation}

\section{Results}

\subsection{The $^{11}$Be + $^{208}$Pb reaction}

We now apply the coupled-channels formalism presented in the previous 
section to fusion of drip-line nuclei. 
We first consider the fusion reaction 
$^{11}$Be + $^{208}$Pb, where the projectile is generally regarded 
as a good example of a single neutron ``halo''.  
In a pure single-particle picture, the last neutron in $^{11}$Be
occupies the $2s_{1/2}$ state, bound by about 500 keV.
The strong concentration of strength at the continuum threshold evidenced in
break-up reactions~\cite{N97} has been mainly ascribed to the promotion of 
this last bound neutron to the continuum of $p_{3/2}$ and $p_{1/2}$ states
at energy $E_c$ via the dipole field.~\cite{N97,EBB95} 
Since the presence of the first excited $1p_{1/2}$ 
state (still bound by about 180 keV) may perturb the transition to the 
corresponding $p_{1/2}$ states in the continuum,~\cite{DLV99} we prefer 
here to consider only the contribution to the $p_{3/2}$ states.
The couplings to the $p_{1/2}$ states are certainly not negligible, 
but we expect that they will simply further enhance the effects on the fusion 
cross section and therefore will not alter our qualitative conclusions.
The initial $2s_{1/2}$ bound state and the continuum $p_{3/2}$ states 
are generated by Woods-Saxon single-particle potentials whose depths
have been adjusted to reproduce the correct binding energies for the
$1p_{3/2}$ and $2s_{1/2}$ bound states.  In particular, one needs for the
latter case a potential which is much deeper than the ``standard'' one.
As we mentioned in the previous section, the form factor $F(r;E_c)$ as 
a function of the internuclear separation $r$
and of the energy $E_c$ in the continuum is then obtained by folding the
corresponding transition density with the external field generated by the
target.  
In addition to the Coulomb field, a Woods-Saxon nucleon-nucleus potential  
is used, with parameters of $R=r_0 A_{_T}^{1/3}$, $r_0=1.27$ fm, 
$a=0.67$ fm, $V=(-51+33~(N-Z)/A)$ MeV, and $V_{ls}=-0.44V$.  

\begin{figure}[p]
\begin{center}
\leavevmode
\parbox{0.8\textwidth}
{\psfig{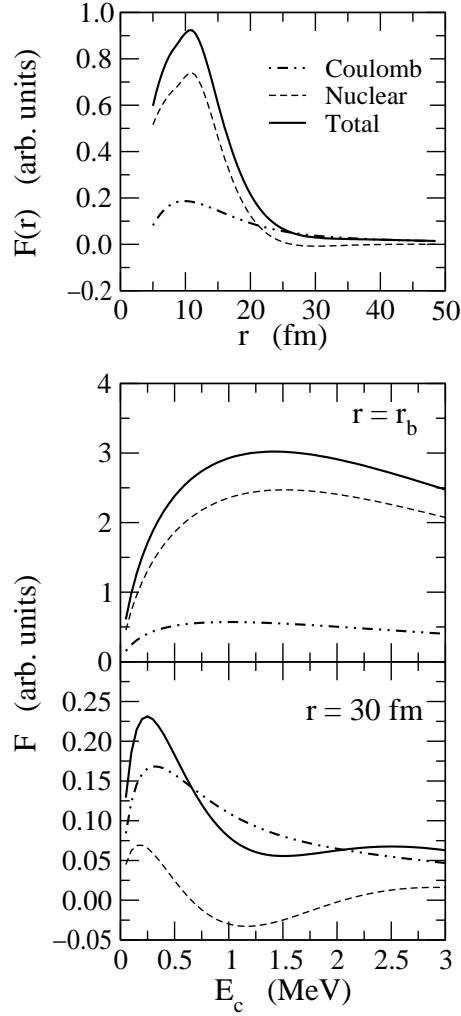}}
\end{center}
\protect\caption{
Coupling form factor $F(r;E_c)$ associated with the 
dipole transition in
$^{11}$Be from the neutron bound state $2s_{1/2}$ ($E_b$ = $-$500 keV)
to the continuum state $p_{3/2}$ with energy $E_c$.  
In (a) the Coulomb (the dash-dotted line), nuclear (the dashed) and 
total (the solid) 
form factors are shown as a function of $r$ at the continuum energy 
$E_c=0.9$ MeV.  In (b) and (c) the form factors are shown as a function
of the energy $E_c$ in the continuum for $r=r_{barrier}$=11.6 fm and $r=30$ 
fm, respectively. 
}
\end{figure}

Selected cuts of the form factor $F(r;E_c)$ are shown in Figure 1.  The 
individual contribution arising from the Coulomb and nuclear interactions
are shown separately. In Fig. 1 (a), 
we display the form factor as a function of $r$ for a fixed value of the 
energy in the continuum ($E_c$ being 0.9 MeV).  
Note the long tail of the nuclear contribution
as a consequence of the large radial extension of the weakly-bound wave 
function, resulting in the predominance of the nuclear form factor up
to the unusual distance of about 25 fm.  The same reason gives rise to a 
deviation of the 
Coulomb part from the asymptotic behaviour $1/r^2$. 
Note also, due to the negative $E1$ effective charge of the neutron
excitation, the constructive interference of the nuclear and Coulomb parts.
In figs. 1 (b) and 1 (c), we 
show, instead, the energy dependence of the form factor at fixed value
of $r$.  While at large $r$ the curves are peaked at very low energies,
reflecting the corresponding $B(E1)$, at smaller distances around the
barrier, for values which are more relevant to the fusion process, the 
peaks of the distributions move to higher energies, in particular for the
nuclear part. 
Hence, the sudden approximation employed in Refs. 7 and 8, where 
the energy of continuum states $E_c$ was all set to be zero, is not 
justified. 

In order to perform the coupled-channel calculation, the distribution of 
continuum states is discretized in bins of energy, associating
to each bin the form factor corresponding to its central energy. We have
considered the continuum distribution up to 2 MeV, with a step of 200 keV.
In this way, the calculations are performed with 10 effective
excited channels.  
We have checked the convergence concerning the maximum energy of the 
continuum states included in the calculation, and found that it is rather 
slow. We have, however, found that our main 
conclusions again do not qualitatively change provided that the maximum 
excitation energy of the continuum states is at least 2 MeV, 
as considered in this work. 
The ion-ion potential is assumed to have a Woods-Saxon 
form with parameters $V_0= -$152.5 MeV, $r_0$=1.1 fm and 
$a=$0.63 fm, a set that leads to the same barrier height as the 
Aky\"uz-Winther potential. 

\begin{figure}[t]
\begin{center}
\leavevmode
\parbox{0.8\textwidth}
{\psfig{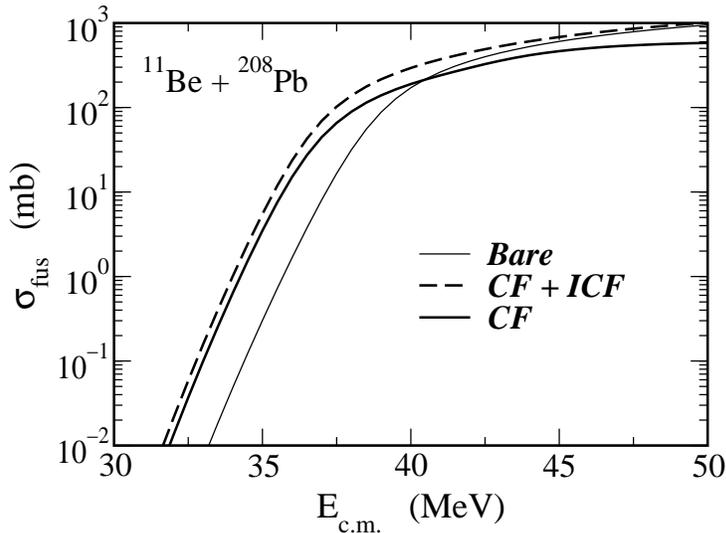}}
\end{center}
\protect\caption{
Fusion cross section for the reaction $^{11}$Be + $^{208}$Pb 
as a function of the bombarding energy in the center of mass frame. 
The thin solid curve shows the results of the one-dimensional barrier 
penetration as a reference.  
The solid and the dashed lines are solutions of the coupled-channels 
equations for the complete fusion and the complete plus incomplete fusion, 
respectively. }
\end{figure}

In the previous theoretical studies of fusion of drip-line 
nuclei,~\cite{MPCD92,TKS93,DV94} the excitations to the ``soft-dipole mode''  
of the projectile were also included in the calculations and the effects of 
the break-up were thus somewhat perturbed. 
In the present calculations, in order to isolate the genuine effect of 
the break-up process, 
we include only the continuum states in the coupling scheme, 
neglecting continuum-continuum coupling as well as other inelastic 
channels such as bound excited states 
in either reaction partner. 
For the same reason, we do not take into account static 
modifications on the ion-ion potential which may arise from the halo 
properties of the projectile.~\cite{TS91} 
Our calculations, therefore, only attempt to give qualitative indications.  
They will, however, still reveal interesting aspects of effects of couplings 
of the ground to continuum states on subbarrier fusion. 

Figure 2 shows the results of our calculations. 
The solid line represents the cross section of complete fusion, 
leading to $^{219}$Rn, while the 
dashed line denotes the sum of the complete and incomplete fusion 
cross sections. Also shown for comparison, by the thin solid line, is 
the cross section in the absence of the couplings to the continuum
states. 
One can see that 
they enhance the fusion cross sections 
at energies below the barrier over the predictions of 
a one-dimensional barrier penetration model.
Note that this is the case not only for the 
total (complete plus incomplete) fusion probability, but also for the 
complete fusion in the entrance channel.  
This finding 
supports the results of the original calculation performed 
in Ref. 9. 
As it has been emphasized there, accounting properly for the dynamical 
effects of the coupling in the classically forbidden region is essential
to arrive at this conclusion.

The situation is completely reversed at energies above the 
barrier. Fusion in the break-up channel becomes more
important and dominates at the expense of the complete fusion. 
As a consequence, the cross sections for complete fusion are 
hindered when compared with the no-coupling case. 

\subsection{The $^{4,6}$He + $^{238}$U reactions}

We next consider the fusion reactions $^{4,6}$He + $^{238}$U. 
This is partly 
motivated by the recent measurement by the SACLAY group.~\cite{T00}
They compared the fusion cross sections of these two systems and 
concluded that the cross sections for the $^6$He projectile were  
systematically larger than those for the $^4$He projectile, both below and 
above the Coulomb barrier. 
One might think that this conclusion is inconsistent with 
ours discussed in the previous subsection as well as with the experimental 
observation of the Canberra group~\cite{D99} that the break-up 
hinders complete fusion cross sections at energies above the 
Coulomb barrier. 
However, the situation is somewhat more complicated
since the experimental data of the SACLAY 
group~\cite{T00} likely include both the complete and incomplete fusion 
cross sections, and also since the halo structure of $^6$He may alter 
significantly the static fusion potential. 
We therefore decided to carry out coupled-channels calculations for 
these systems 
in order to clarify the role played by the break-up of the 
$^6$He projectile. 

To this end, we describe the structure of $^6$He using the di-neutron 
cluster model as in 
Ref. 23. 
The potential between the di-neutron and $\alpha$ particle is determined 
so as to reproduce the correct binding energy of $^6$He. 
The ground state is assumed to be the 2$s$ state in this potential. 
In the coupled-channels calculations, we include continuum states with 
$L=1$ and $L=2$ spins. To construct these wave functions, we slightly 
adjust the depth of the potential for $L=2$ so as to reproduce the 
correct resonance state at 1 MeV. We include the continuum states up to 
$E_c$ = 5 MeV with energy step of 0.25 MeV. 
As for the bare potential, we use the Woods-Saxon potential with 
$V_0=-$174.05 MeV, $r_0$=1.01 fm, and $a$=0.67 fm for the 
$^{4}$He + $^{238}$U system. The potential for the 
$^{6}$He + $^{238}$U system is constructed with the folding procedure 
using the same di-neutron cluster model. 
For simplicity, we neglect the deformation effects of the $^{238}$U 
target which would influence the both reactions in a similar way, 
although one definitely has to include them when one compares 
theoretical fusion cross sections with the experimental data. 

\begin{figure}[t]
\begin{center}
\leavevmode
\parbox{0.8\textwidth}
{\psfig{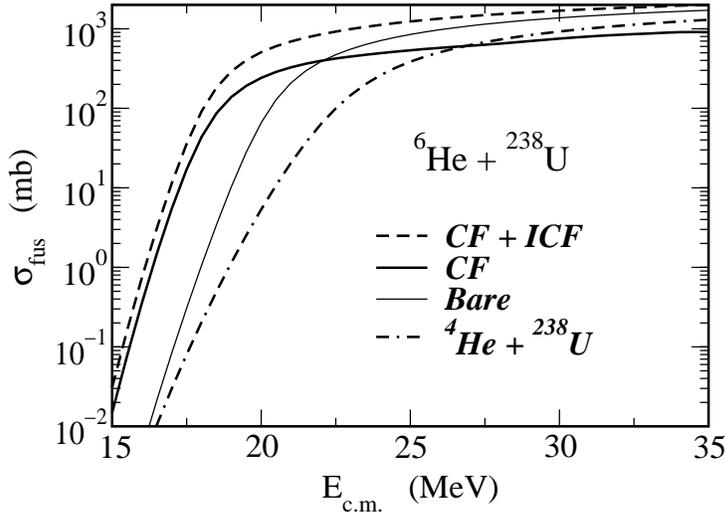}}
\end{center}
\protect\caption{
Same as fig.2, but for the $^{6}$He + $^{238}$U reaction. 
The meaning of each line is the same as in fig.2 except for 
the dot-dashed line, which denotes fusion cross sections for the 
$^{4}$He + $^{238}$U system. }
\end{figure}

Figure 3 shows the theoretical fusion cross sections for the 
$^{6}$He + $^{238}$U reaction and their comparison with 
the $^{4}$He + $^{238}$U system. The relation between the complete fusion 
cross sections (the solid line) and the bare cross sections (the thin solid 
line) within the $^{6}$He + $^{238}$U system is qualitatively the same as that 
in the $^{9}$Be + $^{208}$Pb system discussed in the previous subsection; 
the coupling to the break-up channels enhances the complete fusion cross 
sections at energies below the barrier and hinders them at above the 
barrier energies. 
The cross sections for the $^4$He projectile are denoted by the 
dot-dashed line. 
The figure shows that 
the complete fusion cross sections for the 
$^{6}$He + $^{238}$U reaction are even smaller than the cross sections for 
the $^{4}$He projectile at energies above the Coulomb barrier. 
Notice that the sum of the 
complete and incomplete fusion cross sections for the $^6$He projectile is 
always larger than the cross sections for the $^4$He projectile, that 
reminisces the experimental observation by the SACLAY group.~\cite{T00}   
It would thus be interesting, but perhaps experimentally 
challenging, to isolate the contribution 
from the complete fusion in the 
$^{6}$He + $^{238}$U system. 

\section{Summary}

We have performed exact coupled-channels calculations for 
weakly-bound systems using 
realistic coupling form factors to discuss effects of break-up on 
subbarrier fusion reactions. 
As examples, we have considered the fusion of
$^{11}$Be with a $^{208}$Pb target as well as 
the $^{4,6}$He + $^{238}$U fusion reactions. 
We found that the coupling to break-up channels enhances cross sections 
for the complete fusion at energies below the Coulomb barrier, while it 
reduces them at energies above. 

Very recently, a complete fusion excitation function was measured for 
the $^9$Be + $^{208}$Pb reaction at near-barrier energies by 
Dasgupta {\it et al}.~\cite{D99} They showed that cross sections for 
complete fusion are considerably smaller at above-barrier energies 
compared with a theoretical calculation that reproduces the total 
fusion cross section. 
Also, the fusion cross sections for the 
$^{4,6}$He + $^{238}$U systems measured by the SACLAY group~\cite{T00} 
seem to indicate that the break-up effects enhance fusion cross sections 
at energies below the barrier. 
These two recent experiments are in general agreement with our results. 

\section*{Acknowledgments}
We thank C.H. Dasso and S.M. Lenzi for discussions. 
The work of K.H. was supported by 
the U.S. Dept. of Energy under Grant No. 
DE-FG03-97ER4014.

\end{document}